# Dynamical Networks for Smog Pattern Analysis

Linqi Zong[1], Xinyi Gong[1], Jia Zhu[1,2*]

[1]*College of Engineering and Applied Sciences, Nanjing University, Jiangsu 210093, P.R. China*

[2]*Aerosol Bioeffects and Health Research Center, Nanjing University, Jiangsu 210093, P.R. China*

Smog, as a form of air pollution, poses as a serious problem to the environment, health, and economy of the world[1-4] . Previous studies on smog mostly focused on the components and the effects of smog [5-10]. However, as the smog happens with increased frequency and duration, the smog pattern which is critical for smog forecast and control, is rarely investigated, mainly due to the complexity of the components, the causes, and the spreading processes of smog. Here we report the first analysis on smog pattern applying the model of dynamical networks with spontaneous recovery. We show that many phenomena such as the sudden outbreak and dissipation of smog and the long duration smog can be revealed with the mathematical mechanism under a random walk simulation. We present real-world air quality index data in accord with the predictions of the model. Also we found that compared to external causes such as pollution spreading from nearby, internal causes such as industrial pollution and vehicle emission generated on-site are more critical to the health of the whole network. Most strikingly, we demonstrate that spreading out the pollution sources by moving them to nearby suburban areas, a common practice for smog control, will degrade the overall health of system under certain conditions.

**As we all know, smog poses as a serious challenge to the climate, environment, health, and economy of the world. Here we report the first analysis on smog pattern applying the model of dynamical networks with spontaneous recovery, which reveals the mechanisms of many phenomena such as the sudden outbreak and dissipation of smog and the long duration smog. The real-world air quality index data is also presented in accord with the predictions of the model. According to our study, internal causes such as industrial pollution and vehicle emission generated on-site are more critical to the health of the whole network, compared to external causes such as pollution spreading from nearby. Therefore, direct decrease of the internal causes is a more efficient way to reduce the smog.  The most striking aspect of this work is that we found that spreading out the pollution sources by moving them to nearby suburban areas, a common practice for addressing the smog issue, will not always improve the health of the network.  In some cases, it will actually degrade the health of the whole networks.  Therefore, it provides a theoretical guideline for effective smog control.**

Smog has many detrimental components, such as metallic elements, ionic species, and carbonaceous contents [11]. More frequent smoggy weather is likely to induce extreme climate incidents, hinder air, water and land traffic due to low visibility, affect lung functions and the human immune system, and lead to more deaths and severe chronic diseases. In the recent decades, it struck many metropolises, most notably Los Angeles[12, 13], London[14] and Beijing[15]. Therefore, addressing air pollution and reducing smog is both urgent and necessary. Most of previous works have focused on the chemistry, physiology of components of smog. However, few study investigated the mechanism behind smog pattern, which is crucial for macroscopic control of smog. With the emergence of Erdoˊs Renyi network[16-18], "small-world" networks[19] and scale-free networks[20], dynamical networks have become a powerful framework revealing the mechanism behind many real-world phenomena, such as epidemic infection in biology or collective financial failure in finance[21]. Here for the first time, we study the smog pattern by applying the model of dynamical networks with spontaneous recovery. To construct the networks corresponding to the state without and with smog, we define each node as healthy or damaged correspondingly. The state of an appropriate sized area will be evaluated as the fraction of healthy nodes in the network. The causes of damaged state can be internal or external. Internal causes are pollution generated inside one node, such as photochemical pollution, vehicle emissions, and seasonal human activity like burning coal for heat in cold weather, while external causes are taking-in pollution, such as moving-in of polluted industries and the spread of pollution through wind from nearby nodes. Both internal and external causes can be controlled at different levels with different costs. Apart from the possibility of being damaged, the nodes in the network also have chance to recover from damage through different mechanisms, such as precipitation or air purification.

We can develop the network with three fundamental assumptions: 1. The possibility of damage by internal causes during time interval dt for each node is p dt; 2. If a node has m or less than m healthy neighbors, that node will be in danger of suffering damage by external causes. The possibility of damage by external causes during time interval dt for that node is r dt; 3. After a time period of $\tau \neq 0$, internally damaged nodes can recover to be in healthy state. Similarly, after $\tau' \neq 0$, nodes can recover from external damage. We set $\tau$ as 3 and $\tau'$ as 2.

We can use z and a=1- z to represent the fraction of healthy and damage nodes in the network. In further study, we find parameter p can be substituted by a more convenient parameter p*, which represents the faction of internally damaged nodes and $p^* \equiv 1-e^{-p\tau}$. We use mean-field theory (MFT) for analytical results. For arbitrary networks with neighbor distribution of fk, the fraction of damaged nodes a is described as

$$a(r,p^*) = p^* + r(1-p^*) \sum_k f_k \sum_{j=0}^m \binom{k}{k-j} a^{k-j}(1-a)^j$$

.

The health of system (z) can be determined by internal causes (p*) and external causes (r). We find the relationship can be described as curved surface (figure 1a) in the 3D graph. The diagram can be divided into three parts: high z phase region, low z phase region and multi-value region. Generally speaking, z tends to be larger (smaller) when p* and r are smaller (larger), which is intuitive: if both internal and external causes of smog are minimized, there should be no air pollution problem in the network. As shown in figure 1a and 1b, there is a region where a certain set of p* and r has three possible z values, two of which is physical stable solutions while the other is unstable. In this multi-value region where a certain set of p* and r has multiple z values, the network is unstable. It is obvious in the diagram, when p* is close to 1 or 0 the system has less dependence on r, and depends mainly on p*. Also as we can see in figure 1c and 1d, when p* is less than 0.35, the system exhibits a strong hysteresis behavior, and the health of system (z) is very stable. Therefore, internal causes of damage (p*) are more crucial to determine the health of networks, compared to external ones.

According to the basic assumptions, the fraction of externally damaged nodes among those nodes which have m (or less than m) healthy neighbors should be r at random time t. However, due to probabilistic nature, such as wind direction and wind speed which will change over time, the fraction is not exactly r, but fluctuates around r. In addition, in real-world situations, the influential factors won't always remain the same but vary within reasonable ranges during a certain period of time. For example, during the winter time Air is more likely to be polluted internally if the need of burning coal increases as it gets colder. Thus, it is appropriate to use a random walk of r and p* to simulate the process of factors' influence varying during some time interval. Total range of p* and r in this process is set as r ∈ (0.7, 0.85) and p*∈ (0.2, 0.33) and the step size of p* and r is controlled within 0.01 in order to simulate the fluctuation. The trajectory of the random walk is presented in figure 2b and the corresponding z-time graph is shown in figure 2a.

After detailed analysis, we find two forms of change between high and low z states, which is referred as flash crash and phase change. In figure 2a, the value of z drops at point 1 (t=467) then rapidly recovers. At point 2 (t=1808), the system turns from a relatively healthy state into a rather damaged stable state. More specifically, figure 2b, 2c, 2d and 2e show that flash crash and phase change both experience the trajectory crosses the upper spinodal (the boundary between multi-value region and low z phase region) in the phase diagram. The difference between them is the time interval that p* and r stay in the low z phase region. In phase change, after crossing spinodal, the trajectory stays more that 10 time steps before it could return to the multi-value region. In flash crash, the trajectory leaves the multi-value region for a very short time, and then returns to multi-value region without triggering the phase transition. For real-life situations, if the area is already not clean (in the multi-value region), then a simple event like burning coal would push the whole area into flash crash, which means a massive outburst of smog. Furthermore, if that event continues for a rather long time, the state of being in smog could be stable and hard to alter. The same mechanism also applies when we discuss recoveries from damaged state which brings guidance of controlling smog in a macroscopic way. When the area is in a stable state with smog, fast reduction of the pollution source is not enough unless pollution source is contained for a reasonably long period.

In comparison with the real-world situation, we collect the air quality index data of the City of Shijiazhuang from China meteorological administration. Figure 3d shows the index switches back and forth between high and low levels, resembling the flash crash and phase change of the z-t diagram in our simulation. It indicates that our model and simulation has a plausible fit for the behavior in real environmental conditions.

In the discussion above we assume all nodes in the network are uniform. However, in reality the nodes are not uniform, as different places have different level of pollutions (therefore different p values). For example, some big cities might have higher pollution intensity than the places around it because of its large population, heavy industry and traffic. In the following part, we assume that the internal cause of one node (the special node) p is 60% higher than the nodes around it (Fig. 4a). Several different measures can be taken to address the air pollution. One way is "direct reduction", i.e. directly reducing its internal causes, by for example more strict traffic restriction or pollution control(Fig. 4a). Another way to reduce the internal causes of special node is "move-out", for example by moving the heavy pollution industries such as steel mills, coal-fired power plants out to the nearby nodes. In this case, p value of the special node is decreased, while p values of the nodes around it (defined as around nodes) are increased (Fig. 4c). We can find that z value can be increased in "direct reduction" and "move-out", which means that both ways can help to improve the health of the special node (Fig.4b, d). It is found by comparing the improvement of z value in Fig. 4b and d, that in general "direct reduction" is more efficient in improving the health of the special node than "move-out". Also it is noted in the case of "move-out", when r is high and p* is low, z value gets lower after "moving out". In this region, the external causes play a significant role in the failure of the node. However, as p value of the special node is further lowered to be the same as the normal node (Fig .4e), z value of the network gets significantly improved (Fig. 4f) across the whole range of p* and r, even if r is high and p* is low. That is because compared to the increase of external causes, the decrease of internal causes is more effective. As we assume that a node has m or less than m healthy neighbors, that node will be in danger of suffering damage by external causes. In both cases (Fig. 4d, f), the special node will have less than m healthy neighbors. Therefore the further increase of external causes will not make any difference. The decrease of internal causes therefore becomes a dominant factor, which improves the health of network.

Overall, while "direct reduction" is a more efficient way, "move-out" can also help to improve the health of networks. In other words, it is clear that not just the amount of pollution sources matters, the distribution also plays a critical role to determine the general condition of the whole network.

**ACKNOWLEDGMENTS**

This work is jointly supported by the State Key Program for Basic Research of China (No. 2015CB659300), National Natural Science Foundation of China (NSFC No. 11321063) and the Priority Academic Program Development of Jiangsu Higher Education Institutions (PAPD).

Figure 1

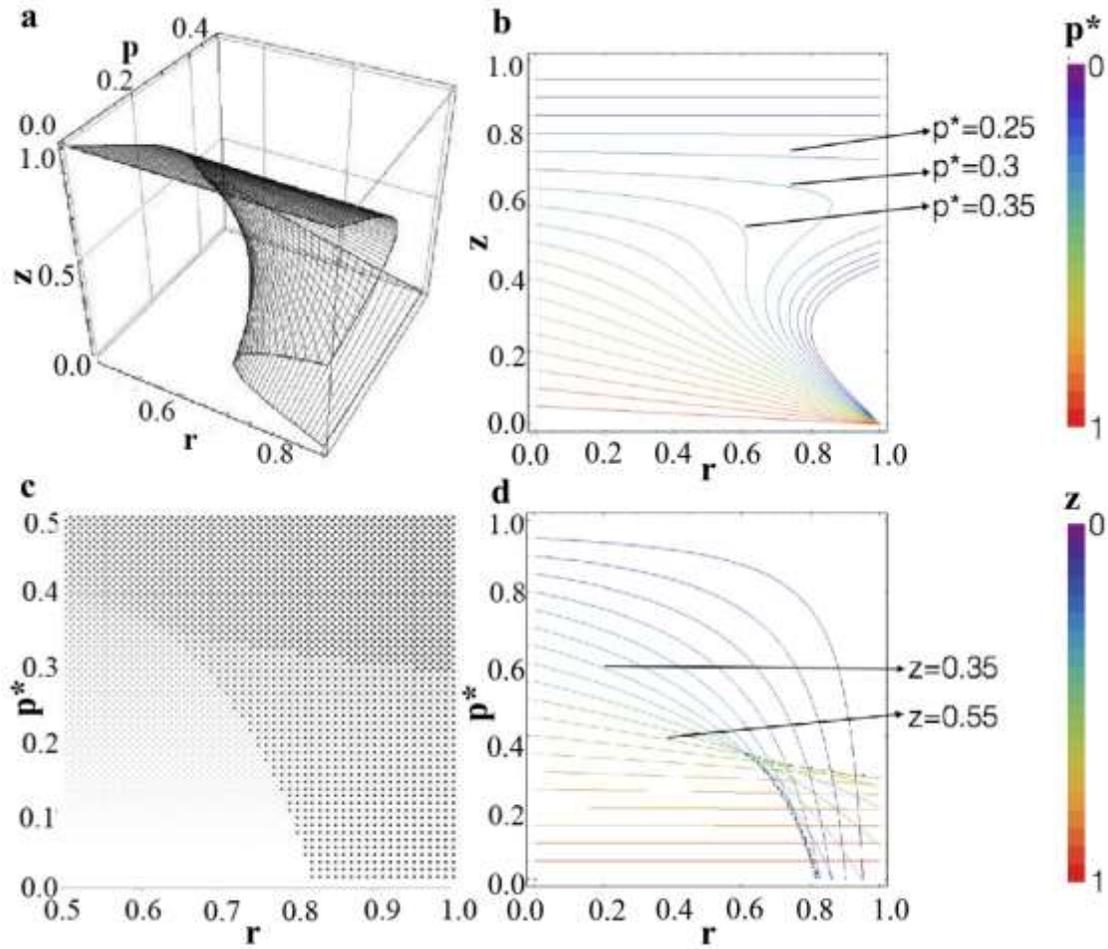

FIG. 1 a) An illustration of the relationship of p*, r and z in 3D graph. The analytical result shows a partly folded curved surface. b) 2D graph shows how p* and r determine the value of z.. c) z-r graph. Each curve represents the relationship between z and r with a certain p* value which is equally spaced from 0 to 1 (step size 0.05). d) p*-r graph. Each curve represents the relationship between p* and r with a certain z value which is equally spaced from 0 to 1 (step size 0.05).

Figure 2

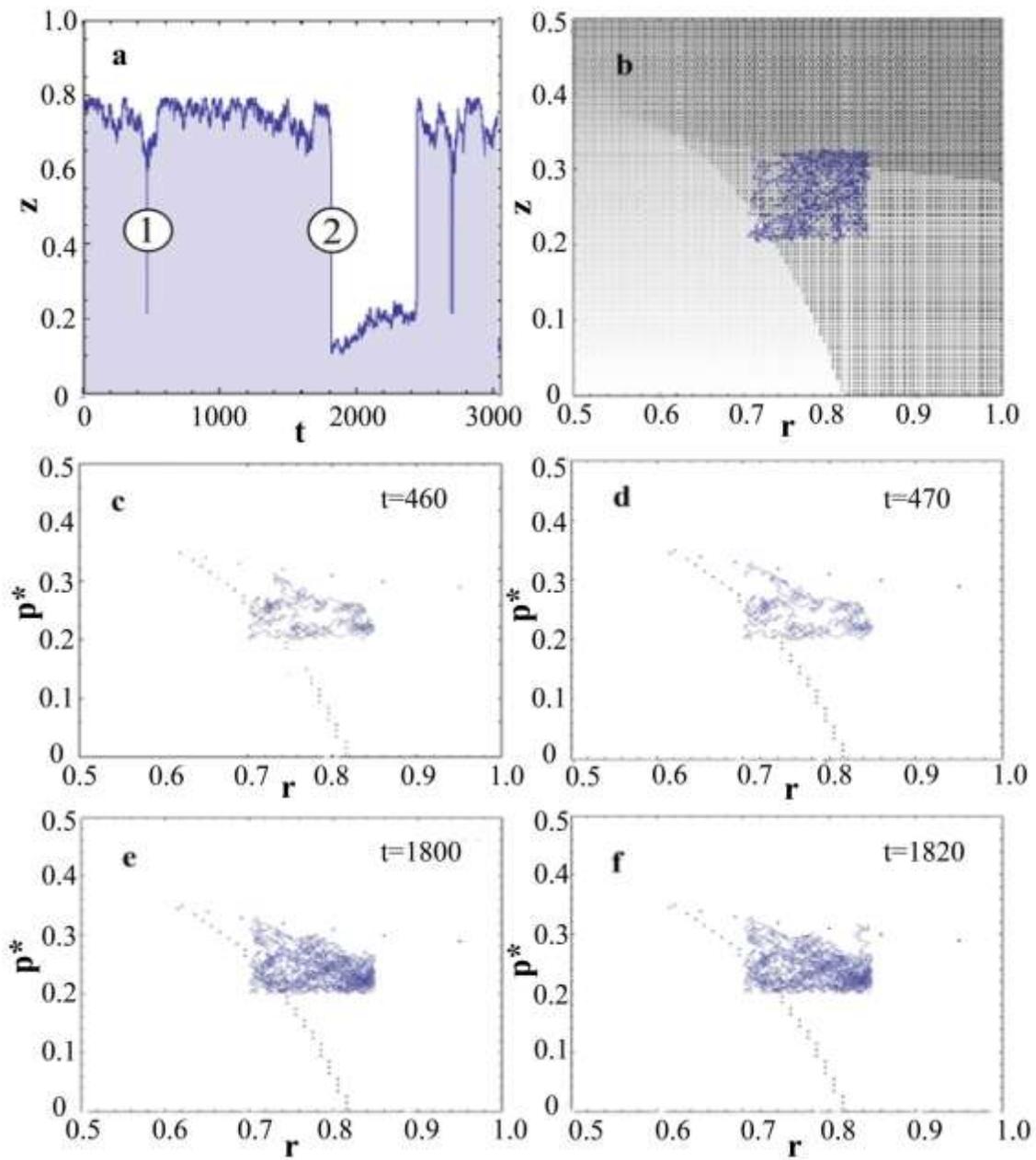

FIG. 2 a)Dynamical switching of the fraction of healthy nodes z with time. Point 1 is flash crash while point 2 is phase change. b), c), d) and e)The trajectory of the random walk with ending time 460, 470, 1800 and 1820.

Figure 3

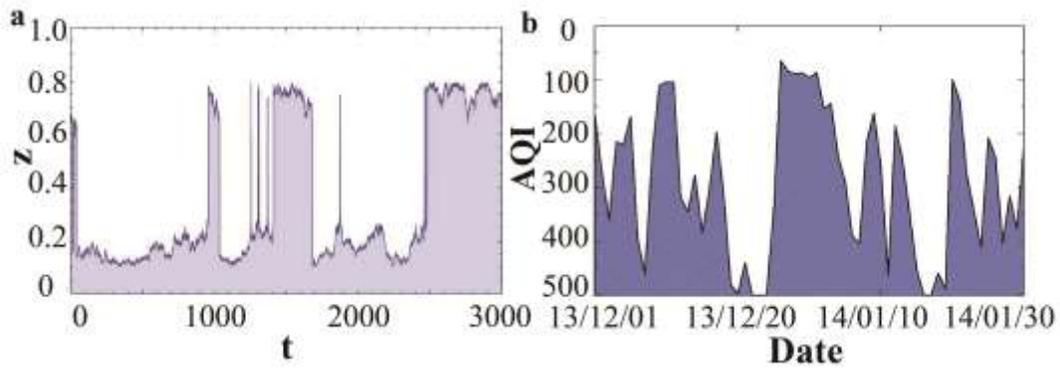

FIG. 3 a) The distribution of z in the random walk corresponding to Fig. 2. b) Air quality index data of Shijiazhuang from December 1th in 2013 to January 30th in 2014 (day-to-day data taken from China meteorological administration) and its distribution.

Figure 4

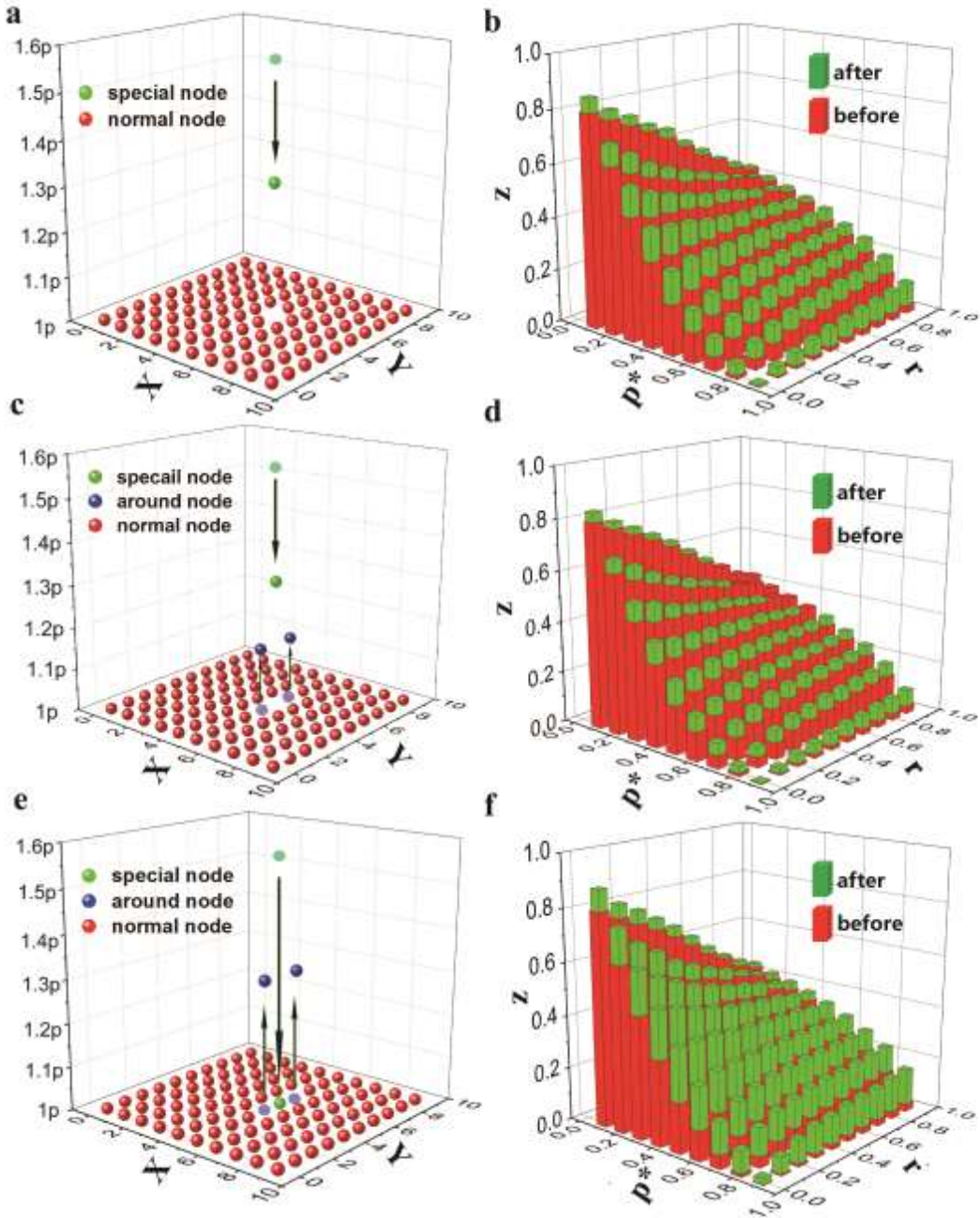

FIG. 4a) Illustration of the nodes in the network by directly reducing the internal causes of the special node ("direct reduction"). b) p*-r-z graph of the result of Fig.4a. c) Illustration of the nodes in the network by "move-out". d) p*-r-z graph of the result of Fig .4c . e) Illustration of the nodes in the network by further "move-out" until the p value of the speical node is the same as the normal node. F) p*-r-z graph of the result of Fig .4e